\documentclass[preprintnumbers,amsmath,amssymb]{revtex4}
\usepackage{graphicx}
\usepackage{dcolumn}
\usepackage{bm}
\begin{document}
\def\oppropto{\mathop{\propto}} 
\def\opsimeq{\mathop{\simeq}}
\def\opoverderline{\mathop{\overline}}
\def\operarrow{\mathop{\longrightarrow}}
\def\opsim{\mathop{\sim}} 

\title{ Random walks and polymers in the presence of quenched disorder }
\author{ C\'ecile Monthus, \\
Service de Physique Th\'{e}orique, CEA Saclay,
91191 Gif-sur-Yvette cedex, France \\ }

 \affiliation{ Conference Proceedings ``Mathematics and Physics", I.H.E.S. ,
 France, November 2005}

\begin{abstract}

\bigskip

After a general introduction to the field, we describe some recent results
 concerning disorder effects on both
`random walk models', where the random walk is a dynamical process generated
by local transition rules, and on `polymer models', where each random walk
 trajectory representing the configuration of a polymer chain is associated
 to a global Boltzmann weight.
For random walk models, we explain, on the specific examples of the Sinai model
and of the trap model, how disorder induces anomalous diffusion, 
aging behaviours and Golosov localization,
and how these properties can be understood via a strong disorder 
renormalization approach.
For polymer models, we discuss the critical properties of various
 delocalization transitions involving random polymers. We first summarize 
some recent progresses in the general theory of random critical points :
thermodynamic observables are not self-averaging at criticality whenever 
disorder is relevant, and this lack of self-averaging is directly
 related to the probability distribution
of pseudo-critical temperatures $T_c(i,L)$ over the ensemble of
samples $(i)$ of size $L$. We describe the results of this analysis
 for the bidimensional wetting and for
the Poland-Scheraga model of DNA denaturation.
 
\end{abstract}
\maketitle

\section{Introduction}
 
 Random walks and diffusion processes have been the subject of
 constant interest in mathematics and in physics
during the last century,  for two main reasons :
 on one hand, they play a central role in probability theory,
and present a large number of very nice mathematical properties;
on the other hand, they naturally appear
 in a great variety of situations in physics and in biology.
We refer the interested reader
to the various reviews that have appeared in 2005 to celebrate
the 100 years of Einstein's theory of Brownian motion,
written either by mathematicians \cite{kahane,legall},
by physicists \cite{derridabrunet,duplantier,majumdar},
or by `biology-oriented' physicists \cite{chowdhury,frey}.
    
In the following, we will be interested into
 the effects of quenched disorder on random walks :
 are the usual properties of random walks stable with respect
 to the presence of some disorder or inhomogeneity ?
 if not, what are the new properties induced by disorder?
 
 To answer these questions,  it is important to distinguish from the very beginning
  what we will call `random walk models',
 where a random walk trajectory is generated by local dynamical rules,
 from  what we will call `polymer models' , where each random walk trajectory
is associated to a global Boltzmann weight.

 \subsection{ Random walk models : trajectories generated by local dynamical rules }
 
 The usual random walk model is defined on a one-dimensional lattice $\{ n \in Z \}$
 by the following local dynamical rule : the particle starts at $n=0$ at time $t=0$,
 and then at each time step, the particle jumps either on the right or on the left
 with probabilities $(1/2,1/2)$.  The position $n(t)$ reached at time $t$
 scales with the well-known diffusive behaviour $n(t) \sim t^{1/2}$.

 Quenched disorder can be introduced in two different ways in the local dynamical rules,
 either in the relative probabilities to jump on the right or the the left,
 or in the time spent on a site before jumping out of it : these two types
 define the two basic `random random walk' models, known respectively
 as the Sinai model and the trap model.

 \subsubsection{ Sinai model : random bias on each site}
 
    The so called `Sinai model'
has interested both the mathematicians since the works
of Solomon \cite{solomon}, Kesten {\it et al.}  \cite{kestenetal}, 
Sinai \cite{sinai}, and the physicists since the works of Alexander
 {\it et al.}\cite{alexander}, Derrida-Pomeau \cite{derridapomeau}.
 There have been many developments in the two communities for the last thirty years: we refer the reader to the recent reviews \cite{zhanshi,zeitouni} for the mathematical side and to the reviews \cite{haus, havlin, jpbreview,revue} for the various physicists approaches. 

 In the Sinai model, each site $n$ of the lattice is characterized by 
 a quenched random variable $\omega_n \in ]0,1[$ representing the probability to jump on
 the right, whereas $(1-\omega_n)$ represents the probability to jump on
 the left. Once these random variables $(\omega_n)$ have been drawn on the full line,
 the random random walk is generated as follows : the particle starts at $n=0$ at time $t=0$,
 and then at each time step, the particle being on site $n$ jumps either on the right or on the left
 with probabilities $(\omega_n,1-\omega_n)$.  
Note that this process naturally appears in various contexts, for instance in the dynamics of a domain wall in the random field Ising chain \cite{rgrfim} or in the unzipping of DNA
 in the presence of an external force \cite{lubenskinelson}.

 \subsubsection{ Trap model : random trapping time on each site}

    Trap models propose a very simple mechanism
 for aging \cite{jpbtraps}, and have been thus
    much studied from this point of view, either in the 
mean field version \cite{jpbtraps, benarousreview,boviertrap},
 or in the one-dimensional
 version \cite{isopi, bertinjp,bertinthese,trapsymmetric,
trapnonlinear,trapreponse}, 
which presents two characteristic time scales for aging,
in contrast with the mean field case. 
    
   In this 1D version of the trap model,
each site $n$ of the lattice is characterized by 
 a quenched random variable $\tau_n$ representing the mean trapping time
 spent at site $n$ before jumping either to the right or to the left
 with probabilities $(1/2,1/2)$. Once these random variables
 $(\tau_n)$ have been drawn on the full line,
 the continuous random walk is defined by the master equation
 \begin{eqnarray} 
\frac{d P_t(n)}{dt} = - \frac{P_t(n)}{\tau_n}
+\frac{P_t(n-1)}{2\tau_{n-1}} +\frac{P_t(n+1)}{2\tau_{n+1}} 
\label{masterdirected1} 
\end{eqnarray} 
Again, this process naturally appears in various physical contexts
 \cite{alexander, jpbreview, bubbledna}.

  \subsubsection{ Disorder  effects :  anomalous diffusion, 
aging and Golosov localization}
  
The random random walks such as the Sinai walk and the trap model
defined above, represent simple dynamical models containing 
quenched disorder, in which many properties such as aging 
that exist in more complex systems 
\cite{crash,dynamicreview,leticialeshouches} can be studied in details.
The first effect of disorder
  is to slow down the dynamics, because disorder
  effectively generates trapping regions from which
 it is difficult to escape.
  For instance, the diffusion
  becomes logarithmic $x(t) \sim (\ln t)^2$ in the symmetric
Sinai model, or can be algebraic $x(t) \sim t^a$
  with some exponent $a$ smaller than $1/2$
in the symmetric trap model, if the mean trapping time diverges.
Besides this anomalous diffusion length scale $x(t)$
that represents the typical distance travelled during time t,
 there exists a second important length scale $y(t)=x_1(t)-x_2(t)$  
that represents the distance 
between two independent particles diffusing in the same
disordered sample from the same initial condition.
In the symmetric Sinai model, as discovered by Golosov,
 the variable $y(t)$ remains finite with probability
 $p=1$ in the limit of infinite time :
 this is the remarkable Golosov localization phenomenon.
 More generally, in the anomalous diffusion phases
of the biased Sinai model and of trap models,
it turns out out that localization still exists,
but is only partial  : the variable $y(t)$ remains finite
 with probability $0<p<1$ in the limit of infinite time.

In Section \ref{randomwalks}, we explain how 
these random walks presenting some localization
 may be studied via a strong disorder analysis, which yields results 
both for one-time observables and for aging properties.

 \subsection{ Polymer models : Boltzmann weight on random walk trajectories } 
 
 In `polymer models', random walk trajectories are not considered
 as the results of a dynamical process, but represent conformations
 of a `polymer' chain, on which one defines a probability measure.
 To see the difference, the most well-known example is
 the Self-Avoiding-Walk (SAW) measure in dimension $d \geq 2$ defined as follows :
 among all the random walks of $N$ steps, the weight is uniform
 for all random walks having no self-crossing, and zero otherwise.
 It is clear that this SAW measure cannot be generated by local dynamical rules.
 
 In the following, we will be interested into phase transitions
of directed polymer models. Here one has to distinguish two types
of models, namely models where the phase transition already exists
in the pure case, and models where the phase transition only exists
in the disorder case. We now introduce one example of each type.

  \subsubsection{ Bidimensional wetting :
 disorder effects on critical points }

  Wetting transitions are in some sense the simplest phase transitions,
since they involve linear systems \cite{mfisher}. 
  Let us consider 
  a one-dimensional random walk of increments 
$z(\alpha+1)-z(\alpha)=\pm 1$,
  starting at $z(0)=0$. The random walk is constrained to remain 
in the upper half plane $z \geq 0$, but gains an adsorption
energy $\epsilon_{\alpha}$
if $z(\alpha)=0$.  More precisely, the model
 is defined by the partition function 
\begin{equation}
Z_{2N}(\beta) = \displaystyle \sum_{RW} 
\exp \left( \beta \displaystyle \sum_{1 \leq \alpha \leq N}
 \epsilon_{\alpha}\delta_{z_{2 \alpha},0}  \right) 
 \label{zwetting}
  \end{equation}
  In the pure case $\epsilon_{\alpha}=\epsilon_0$,
 there exists a phase transition between
  a localized phase at low temperature, characterized
 by an extensive number
  of contacts at $z=0$, and a delocalized phase at low temperature.
  The critical temperature is simply the point where
 $e^{\beta_c \epsilon_0}=2$, i.e.
  where the Boltzmann weight
  of a contact $e^{\beta_c \epsilon_0}$ exactly 
compensate the factor $2$ lost
  at each contact  for entropic reason.
 At $T_c$, the wall $z=0$ is exactly reflexive,
  whereas for $T<T_c$ it is attractive and for $T>T_c$ it is repulsive.
  For this model, as well as in other types of depinning transitions, one is
  interested into the effect of disorder 
  in the contact energies $\epsilon_{\alpha}$ onto the critical properties
  of the phase transition.
  
In Section \ref{wettingPS}, we first summarize 
some recent progresses in the finite size scaling theory of
disordered systems : thermodynamic observables are not self-averaging
at critical points whenever disorder is relevant, and this lack of
self-averageness at criticality is directly related to the
scaling properties of the probability distribution
of pseudo-critical temperatures $T_c(i,L)$ over the ensemble of
samples $(i)$ of size $L$. We then explain how this
 framework is very useful to characterize
 various delocalization transitions involving random polymers,
in particular
the bidimensional wetting transition introduced above,
and the Poland-Scheraga model of DNA denaturation.

 \subsubsection{ Directed polymers in random media : transition towards a disorder-dominated phase }
 
 The model of directed polymer in a $1+d$ random medium 
 is defined by the following partition function
 \begin{equation}
Z_{N}(\beta) = \displaystyle \sum_{RW} 
\exp \left( \beta \displaystyle \sum_{1 \leq \alpha \leq N} \epsilon(\alpha, \vec r(\alpha))  \right) 
\label{directed}
  \end{equation}
over $d-$dimensional random walks $\vec r(\alpha)$, where the
 independent 
random energies $\epsilon(\alpha, \vec r)$ define the random medium.
This model has attracted a lot of attention because it is directly related
to non-equilibrium properties of growth models 
\cite{revuezhang}.
Within the field of disordered systems, it is also very interesting on its own
because it represents a `baby-spin-glass' 
 model \cite{revuezhang,derridaspohn,fisherhuseDP,mezardDP,somendrareview} presenting
a disorder-dominated phase, where the order parameter is an `overlap' :
two copies of the polymer in the same disordered sample 
have an extensive number of contacts. This localization property, that has been
now proven by mathematicians \cite{carmonahu,comets}, is reminiscent
of the Golosov localization of random walks in random media.
From the point of view of anomalous diffusion however, the two types
of models are very different, since the disorder-dominated phase
of directed polymers in random media are super-diffusive
 $x \sim t^{\zeta}$ where $\zeta>1/2$,
because the polymer has more chance to find a best global path
by making larger excursion, whereas random walks in random media
are usually sub-diffusive because disorder generates traps for the dynamics.

In dimensions $d \leq 2$, this disorder-dominated phase
exists at all temperatures, whereas in dimensions $d > 2$, there exists
a phase transition between this disorder-dominated phase at low temperature
and a diffusive phase at high temperature \cite{cookderrida,imbriespencer}.
The critical properties of the phase transition in $d=3$ have been studied
in \cite{derridagolinelli,kimbraymoore}
via standard finite-size scaling on disordered averaged observables.
However, as explained above for depinning transitions, random critical points
can be better understood via the knowledge of probability distribution
of pseudo-critical temperatures $T_c(i,L)$ over the ensemble of
samples $(i)$ of size $L$. We are currently studying
this distribution, and preliminary results show that the critical point is `unconventional',
with different scalings for the shift and the width of this distribution \cite{future}.

  \section{ Random walks in random media}
  
\label{randomwalks}
  
   \subsection{  Sinai walk and diffusion in a Brownian potential }

 \subsubsection{  Continuous and discrete versions of the model }

The continuous version of the Sinai model corresponds to the 
Langevin equation \cite{jpbreview} 
\begin{eqnarray} 
\frac{dx}{dt} = - U' (x(t)) + \eta(t) 
\label{langevin} 
\end{eqnarray}
where $\eta(t)$ represents the usual thermal noise 
\begin{eqnarray} 
< \eta(t) \eta(t ') > = 2 T \delta(t-t ') 
\end{eqnarray} 
and where $U(x)$ is the quenched Brownian potential  
\begin{eqnarray} 
\overline { (U(x)-U(y))^2} = 2 \sigma \vert x-y \vert 
\label{defsigma} 
\end{eqnarray} 
More generally, in the whole paper, thermal averages of observables are denoted by
 $<f>$, whereas disorder averages are denoted by $\overline{f}$.

In the discrete Sinai model on the 1D lattice presented in the Introduction, 
the particle which is on site $i$ has a probability
$\omega_i $ of jumping to the right and a probability $(1-\omega_i) $ of jumping to the left.  The $\omega_i$ are independent random variables in $]0,1[$.  The random walk is recurrent only if $\overline{ \ln \omega_i} = \overline{ \ln (1-\omega_i)}$, 
which corresponds to the absence of bias of the random
potential $U(x)$ (\ref{defsigma}) of the continuous version. 

 \subsubsection{  Logarithmic anomalous diffusion  }

The first important property of the Sinai model
is the logarithmic scaling of the typical displacement $x \sim (\ln t)^2$ \cite{sinai} 
instead of the usual behaviour  $x \sim \sqrt{t}$ of the pure diffusion. 
This very slow dynamics can be understood by the following argument
 argument \cite{jpbreview} :  
the time $t(x)$ necessary to reach the point $x>0$ will be dominated by the Arrhenius factor of $e^{\beta B_x}$ associated to the largest barrier $B_x$
which should be passed by thermal activation to go from the starting point $x=0$ to the point $x$ (This approximation by the Arrhenius factor amounts 
to apply a saddle-point method on the exact expression for the first passage
time).  In a Brownian potential, the typical behaviour of the barrier $B_x \sim \sqrt{x}$ leads to an Arrhenius time $t \sim e^{\beta \sqrt{x}} $, which indeed corresponds to the scaling $x \sim (\ln t)^2$ after inversion. 

 \subsubsection{  Golosov localization  }

The second important property of the Sinai model
is  the Golosov localization \cite{golosovlocali} :
the distance between two
independent particles (i.e. with two independent thermal histories)
which diffuse in the same disordered sample, remains a finite random variable in the limit of infinite time.

This remarkable phenomenon means that there
are sample-dependent areas that concentrate almost all the
probability weight, and that thermal
fluctuations are completely sub-dominant with respect to disorder.

 \subsubsection{  Strong disorder approach  }

In the strong disorder approach of
the Sinai model \cite{rgsinaishort,rgsinailong}, the essential idea is to decompose the process $x_{U, \eta}(t)$, representing the position of the random walk generated by the thermal noise $\eta(t)$ in the random potential Brownian $U(x)$, into a sum of two terms \begin{eqnarray} 
x_{\{U, \eta\}}(t) = m_{\{U\}}(t) + y_{\{U, \eta\}}(t) 
\end{eqnarray} 

$ \bullet $  the process $m_{\{U\}}(t)$ called the ``effective dynamics" 
depends only on the disorder but not on the thermal noise :
 it represents the most probable position of the particle at the moment $t$.
It simply corresponds to the best local minimum of the random potential $U(x)$ that the particle has been able to reach at time $t$.  As the escape over a potential barrier $F$ requires an Arrhenius time of order $t_F=\tau_0 e^{\beta F}$, one can study in detail this effective dynamics by using a strong disorder real space renormalization  which consists in the iterative decimation of the smallest barriers remaining in the system.  One then associates to time $t$ the renormalized landscape in which only barriers larger than the RG scale $\Gamma=T \ln t$
have been kept.   The position $m_{\{U\}}(t) \sim \Gamma^2=(T \ln t)^2$ then corresponds at the bottom of the renormalized valley at scale $\Gamma=T \ln t$ which contains the initial condition at $t=0$.

$ \bullet $ the process $y_{\{U, \eta\}}(t)$ represents the thermal fluctuation 
with respect to effective dynamics.  In the limit of infinite time, it 
remains a finite random variable. This very strong result is the Golosov localization
phenomenon \cite{golosovlocali} :  all the particles which diffuse in the same sample starting from the same starting point with different thermal noises $\eta$ are asymptotically concentrated in the same renormalized valley of minimum $m_{\{U\}}(t)$.  More precisely, if one considers the first corrections at large time, the probability that a particle is not in the valley corresponding to effective dynamics $m_{\{U\}}(t)$ is of order $1/(\ln t)$, in which case the particle is at a distance of order $(\ln t)^2$ from $m_{U}(t)$.  These events are thus rare (their probability tends towards zero to large time) but they nevertheless dominate  certain observables, such as the thermal width $\overline{\Delta x^2(t)} \sim \overline { < y^2(t) >} \sim (\ln t)^3$ which diverges.  

As a consequence of the Golosov localization, the distribution of the rescaled 
variable $X=\frac{x_{\{U, \eta\}}}{ (T \ln t)^2}$, with respect to the thermal noise $\eta$ in a given sample is asymptotically a Dirac delta distribution  $\delta(X-M)$ where $M=\frac{m_{\{U\}}(t)}{ (T \ln t)^2}$ is the rescaled variable of the effective dynamics.  To compute the averaged diffusion front over the samples (or 
equivalently over the initial conditions), it is then enough to study the distribution of $M$ over the samples \cite{rgsinaishort,rgsinailong}. This leads to the Kesten law 
\begin{eqnarray} 
P(X) = LT^{-1}_{p \to \vert X \vert} \left[ \frac{1}{p} \left(1-\frac{1}{\cosh \sqrt p} \right) \right ] = \frac{4}{\pi} \sum_{n=0}^{+\infty} \frac{(-1)^n}{2n+1} e^{ - \frac{\pi^2}{4} (2n+1) \vert X \vert} 
\end{eqnarray} 
which is an exact result of mathematicians \cite{kestenlaw,golosovdemi, golosovhitting}.  This example explicitly shows how the strong disorder RG 
allows to obtain asymptotic exact results, and gives confidence in the new results of the method concerning finer properties. 

Similarly, the rescaled variable for the energy 
\begin{eqnarray} 
w=\frac{U(x(0))-U(x(t))}{(T \ln t)} \simeq \frac{U(m(0))-U(m(t))}{(T \ln t)} 
\end{eqnarray} 
is entirely given at large time by the effective dynamics.  
The reduced variable $w$ has the following limit law as $t \to \infty$:  \begin{eqnarray}
 {  D}(w) = \theta(w < 1) \left(4- 2 w-4 e^{-w} \right) + \theta(w \geq 1) \left(2 e - 4 \right) e^{-w} 
 \end{eqnarray} 
This law is continuous, like its derivative at $w=1$, but the second derivative 
 is discontinuous at $w=1$, which can seem surprising at first sight.
  Indeed, for any finite time, the energy distribution is analytic, and it is only in the limit of infinite time that the discontinuity appears for the rescaled variable.  It is interesting to note that in the recent mathematical work \cite{hu} over the return time to the origin after time $t$,
another not-analytical asymptotic distribution for a rescaled variable also
appears. The joint limit distribution of the position $X = \frac{x(t)-x(0)}{(T \ln t)^2}$ and the energy $ w=\frac{U(x(0))-U(x(t))}{(T \ln t)}$
may also be computed in Laplace transform \cite{energysinai}.

\subsubsection{ Aging properties}

The two-time diffusion front $\overline{P(x, t;  x', \vert 0,0)}$ 
presents an aging regime in $({{\ln t}/{\ln t'}})$.  In the rescaled variables  $X={(x/\ln^2t)}$ and $X'={(x'/\ln^2t)}$, the diffusion front is again determined by the effective dynamics.  The RG procedure allows to compute the joint law of the positions $\{m(t), m(t_w)\}$ at two successive times $t \geq t_w$ 
\cite{rgsinailong}.  In particular, this two time diffusion front presents a 
Dirac delta function $\delta(X-X')$, which means that the particle can be trapped in a valley from which it cannot escape between $t'$ and $t$.  The weight $D(t, t_w)$ of this delta function thus represents the probability of having $m(t)=m(t_w)$
\begin{eqnarray} 
D(t, t_w) = \frac{1}{3} \left(\frac{\ln t_w}{\ln t}\right)^2 \left(5-2 e^{ 1-
 \left(\frac{\ln t}{\ln t_w} \right)} \right) 
\label{dttw} 
\end{eqnarray} 

\subsubsection{ Distribution of the thermal packet} 

The asymptotic distribution of the relative position $y = x(t)-m(t)$ with respect
 to the effective dynamics $m(t)$ 
corresponds to the Boltzmann distribution in an infinite Brownian valley \begin{eqnarray}
 P(y) = \left< \frac{e^{ - \beta U_1( \vert y \vert))}} { \int_0^{\infty} dx e^{ - \beta U_1(x)} + \int_0^{\infty} dx e^{-\beta U_2(x)}} \right>_{\{U_1, U_2 \}} 
\end{eqnarray} 
where the average is over two Brownian trajectories ${\{U_1, U_2\}}$ 
forming an infinite valley.  This formulation is equivalent to the Golosov
theorem \cite{golosovlocali}.  The law can be explicitly computed in Laplace transform in terms of Bessel functions \cite{locgolosov}, and
presents in particular the algebraic decay 
\begin{eqnarray} 
P(y) \opsim_{y \to \infty} \frac{1}{y^{3/2}} 
\label{loi3/2} 
\end{eqnarray} 
This can be understood as follows:  whereas  
 the Brownian potential $U(y)$ yields a typical decay  of order $e^{ - \beta \sqrt{\sigma y}} $ for the Boltzmann factor, there are rare configurations which return close to $U \sim 0$ at a long distance $y$, with a probability of order $1/(y^{3/2})$.  As expected, the correlation of two 
independent particles in the same sample computed in \cite{locgolosov}
\begin{eqnarray} 
C(l) = \lim_{t \to \infty} 2 \int_{-\infty}^{+\infty} dx \overline{ \left[ P(x, t|x_0,0) P(x+l, t|x_0,0) \right ]} 
\end{eqnarray} 
presents the same algebraic decay in $1/l^{3/2}$.  

\subsubsection{ Thermal width and rare events}

 The algebraic decay (\ref{loi3/2}) for the limit law of the relative position $y$ implies that the second moment $\overline{<y^2>}$ diverges at infinite time.  To obtain its leading behaviour at large time, it is in fact necessary to take into account the following rare events \cite{rgsinailong}:  (a) a renormalized valley can have two minima which are almost degenerated in energy;  (b) two neighbouring barriers can be almost degenerated;  (c) 
a barrier can be near the decimation threshold $(\Gamma+\epsilon)$.
 These rare events appear with a weak probability of order $1/\Gamma$, but they give rise to a splitting of the thermal packet into two sub-packets, separated by a long distance of order $\Gamma^2$.  As a consequence, these 
rare events dominate the thermal width \cite{rgsinailong}
\begin{eqnarray} 
\overline{<x^2(t)>-<x(t)>^2} \oppropto_{t \to \infty} \frac{T}{\Gamma} (\Gamma^2)^2 = T (T \ln t)^3
 \end{eqnarray} 
 This behaviour of the thermal width has been measured numerically \cite{chave}.

\subsubsection{ Discussion } 

The strong disorder approach of the Sinai model thus allows to 
obtain many explicit results, and some of them have  
 now been confirmed by mathematicians, in particular 
the results concerning the weight of the singular part of the two time diffusion front \cite{dembo}, the statistics of the returns at the origin of effective dynamics \cite{cheliotis}, or properties of eigenvalues and eigenvectors
of the Fokker-Planck operator \cite{alex}.

As a final remark, let us briefly explain how the strong
disorder approach is related to general theory of slow dynamics
 based on the idea of metastable states 
 \cite{biroli,kurchanfermion,boviermeta,boviermeta2}.
The idea is to decompose the dynamics into two parts.
 There are on the one hand fast degrees of freedom, which convergence quickly towards a local quasi-equilibrium :
 they correspond to the `` metastable states ". On the other hand, there is a slow
out-of-equilibrium dynamics which corresponds to the evolution of the metastable states.  In this language, the strong disorder description 
of the Sinai random walk can be reformulated as follows: 

 $ \bullet $ the metastable states at time $t$ are the valleys of the 
renormalized landscape at scale $\Gamma = T \ln t$:  indeed, the walkers who were at $t=0$ inside this valley have not been able to escape from this valley
before time $t$.  

$ \bullet $ In each renormalized valley, there is a quasi-equilibrium described by a Boltzmann distribution inside the valley.  

$\bullet$ the slow dynamics corresponds to the evolution of the 
renormalized landscape with the scale $\Gamma = T \ln t$:  some metastable states disappear and are absorbed by a neighbour.

    \subsection{ Biased Sinai walk and related directed trap model }

\subsubsection{Anomalous diffusion phase}

 The introduction of a constant force $F_0$ into the Langevin equation (\ref{langevin}) of the Sinai model is very natural.  This biased
model has also interested mathematicians and physicists for a long time, because it presents a series of dynamic phase transitions \cite{kestenetal, derridapomeau, jpbreview} in terms of the dimensionless parameter $\mu=F_0 T/\sigma$.  In particular, there exists an anomalous diffusion phase
for $0<\mu<1$, which is characterized by the asymptotic behaviour 
 \begin{eqnarray} 
\overline{ < x(t) >} \opsimeq_{t \to \infty} t^{\mu} 
\end{eqnarray} 
whereas for $\mu>1$, the velocity becomes finite:  $\overline{ < x(t) >} \sim V(\mu) t$ with $V(\mu)=F_0 (1-1/\mu)$.  
In the anomalous diffusion phase, the exact diffusion front is given in
terms of L\'evy stable distributions \cite{kestenetal,jpbreview,hushiyor}

\subsubsection{Related directed trap model}

It has been proposed for a long time \cite{feigelman, jpbreview,jpbreview2} 
that the biased Sinai model should be asymptotically equivalent 
to a directed trap model defined by the master equation
 \begin{eqnarray} 
\frac{d P_t(n)}{dt} = - \frac{P_t(n)}{\tau_n}+\frac{P_t(n-1)}{\tau_{n-1}} 
\label{masterdirected} 
\end{eqnarray} 
in which the $\tau_n$ are independent random variables 
distributed with the algebraic law
 \begin{eqnarray} 
q(\tau) \opsimeq_{\tau \to \infty} \frac{\mu}{ \tau^{1+\mu}} 
\label{lawtrap} 
\end{eqnarray} 
The anomalous diffusion phase $0<\mu<1$ then corresponds to
the case where the averaged trapping time is infinite.

The directed character of this trap model allows to obtain many exact results,
since the particle visit sites only once in a fixed order, from left to right.
In particular, the diffusion front can be 
expressed in terms of L\'evy stable laws \cite{jpbreview}.
The thermal width has been exactly computed in \cite{aslangul}
\begin{eqnarray}
\overline { < \Delta n^2(t) >}  \equiv  
 \overline { \sum_{n=0}^{+\infty} n^2 P_t(n) -
[ \sum_{n=0}^{+\infty} n P_t(n) ]^2}
= \frac{1}{\Gamma (2 \mu)} \left( \frac{ \sin \pi \mu}{\pi \mu} \right)^3 I(\mu) t^{2 \mu} 
\label{aslanguleq}
\end{eqnarray}
where $I(\mu)$ is some explicit integral \cite{aslangul}.
The result (\ref{aslanguleq}) shows that the the thermal
packet is spread over a length of order $t^{\mu}$.
On the other hand, the infinite-time limit
of the localization parameter
for $k=2$ computed in \cite{compte}
\begin{eqnarray}
Y_2 (\mu) && \equiv \lim_{t \to \infty} 
\sum_{n=0}^{+\infty} \overline{ [P_t(n)]^2} = \int_{-\pi}^{+ \pi} \frac{ d\theta}{2 \pi}
\  \frac{e^{i \theta \mu}-e^{i \theta}}{ 1- e^{i \theta (\mu+1)}}
\label{y2exact}
\end{eqnarray}
shows that $Y_2$ is finite in the full phase $0 \leq \mu<1$
and vanishes in the limit $\mu=1$. 
How can this property coexist with the result (\ref{aslanguleq})
for the thermal width ? The numerical
simulations of \cite{compte}
show that for a single sample at fixed $t$, the probability distribution 
$P_t(n)$ is made out of a few sharp peaks that have a finite weight
but that are at a distance of order $t^{\mu}$. This explains why
at the same time, there is a finite probability to find two particles
at the same site at infinite time, even if the thermal width 
diverges as $t^{2 \mu}$ at large time.

 \subsubsection{ Strong disorder approach} 

The strong disorder RG presented in the previous section 
for the symmetric Sinai model can be extended to the biased case, but the
obtained results are exact in the limit of infinite time $t \to \infty$ only 
if the bias is very small $\mu \to 0$ \cite{rgsinailong}:
for instance, the RG yields an exponential diffusion front for the rescaled
variable $X = \frac{x(t)}{t^{\mu}}$ that coincide with the exact result involving
a L\'evy distribution \cite{kestenetal, jpbreview} only in the limit $\mu \to 0$.  The reason why the effective dynamics is not exact any more when $\mu$ is finite, is that the distribution of the barriers against the bias
converges towards an exponential distribution of finite width proportional to $1/\mu$.  This shows that the localization of the full thermal packet in a single renormalized valley at large time, which is valid in the limit $\mu \to 0$, is not exact any more for finite $\mu$.  
It is thus necessary to generalize the strong disorder RG approach to include the spreading of the thermal packet into several renormalized valleys.
This can be done as explained in
\cite{trapdirected}, and 
from the description of the diffusion front in each sample, one can compute
exact series expansion in $\mu$ for all observables.  
In particular, the explicit computations up to order order $\mu^2$
\cite{trapdirected} of the diffusion front for the rescaled variable $X=\frac{x}{t^{\mu}}$, of the thermal width 
\begin{eqnarray} 
\lim_{t \to \infty} \frac{\overline { < \Delta x^2 (t) >}} { t^{2 \mu}} = \mu (2 \ln 2) + \mu^2 [ - \frac{\pi^2}{6} + 2 \ln 2 (\ln2 -2+2 \gamma_E) ] +O(\mu^3) \label{widthexact} \end{eqnarray} 
and of the localization parameter
\begin{eqnarray} Y_2 (\mu) = 1 - \mu (2 \ln 2) + \mu^2 (4 \ln 2 \frac{\pi^2}{6}) +O(\mu^3) 
\label{y2exactexpansion} 
\end{eqnarray} 
coincide with the series expansions of the corresponding
results, for the diffusion front
 \cite{kestenetal, jpbreview}, for the thermal width given in Eq (\ref{aslanguleq}) and for the localization parameter 
 given in Eq (\ref{y2exact}).  These comparisons with exact results obtained independently shows that the generalized RG procedure is exact order by order in $\mu$. To compute observables at order $\mu^n$, it is thus enough to consider that the diffusion front is spread over $(1+n)$ traps and to average over the samples with the appropriate measure \cite{trapdirected}.
This approach thus allows to understand how the anomalous diffusion phase $0<\mu<1$
presents at the same time a diverging thermal width
 as $t^{2\mu}$ (\ref{widthexact}) 
together with a finite probability $Y_2(\mu)$ (\ref{y2exactexpansion}) of
finding two particles in the same trap at large time.  

Moreover, the strong disorder approach yields a 
quantitative mapping between the biased Sinai model and the
directed trap model at large times.
All the results for the directed trap model can then 
be translated for the biased Sinai
model, one simply has to replace traps by renormalized valleys.
  In particular, the thermal width has for expansion
 \begin{eqnarray} 
\frac{ \overline { < \Delta x^2(t) >}}{t^{2 \mu}} = \frac { \left(\sigma^2 \beta^3 \right)^{2\mu}} { \sigma^2 \beta^4} \left[ \frac{(2 \ln 2)}{\mu^3} + [ - \frac{\pi^2}{6} + 2 \ln 2 (\ln2 -2-2 \gamma_E) ] \frac{1}{\mu^2} +O(\frac{1}{\mu}) \right ] \label{widthsinaidv} 
\end{eqnarray} 
In conclusion, the anomalous diffusion phase $x \sim t^{\mu}$
 with $0 < \mu<1$ of the biased Sinai model is characterized
 by a localization on several  renormalized
valleys, whose positions and weights can be described sample by sample.  
The generalized strong disorder approach allows to compute all observables
via a systematic series expansions in $\mu$.

    \subsection{ Symmetric trap model   }
    
 The symmetric trap model is defined by
 the master equation (\ref{masterdirected1}),
 where the trapping times
 $\tau_n = e^{\beta E_n}$ are defined in terms of random energies
$E_n$  distributed with the following exponential distribution 
 \begin{eqnarray} 
\rho(E) = \theta(E) \frac{1}{T_g} e^{ - \frac{E}{T_g}} 
\label{rhoe} \end{eqnarray} 
This choice of exponential distribution comes from
the exponential tail of the Gumbel distribution  which represents an important universality class for extreme statistics. 
 The exponential distribution of energies (\ref{rhoe}) translates for the trapping time $\tau = e^{\beta E}$ into the algebraic law 
\begin{eqnarray} 
q(\tau) = \theta(\tau>1) \frac{\mu}{\tau^{1+\mu}} 
\label{qtau} 
\end{eqnarray}
 with exponent
 \begin{eqnarray} 
\mu = \frac{T}{T_g} \label{defmu} 
\end{eqnarray} 
At low temperature $T<T_g$, the average trapping time 
$\int d \tau \tau q(\tau)$ diverges, and this directly leads to aging effects.
    
In the { \it  symmetric} model, each site can be visited several times, 
which leads to an essential change in the strong disorder approach
 \cite{trapsymmetric} with respect to the directed trap model : 
 a trap of the renormalized landscape will be characterized by two important times, namely (i) its trapping time $\tau_i$, which represents the typical time of exit towards its immediate neighbours (ii) its 
escape time, which represents the time needed to reach a deeper trap.

In the limit $\mu \to 0$, the following effective dynamics becomes exact:
  at time $t$, the particle starting from the origin at $t=0$
will be at time $t$ either on the first renormalized trap $M_+$ at distance $l_+$
on the right or on the first renormalized trap $M_-$ at distance $l_-$  on the left.  The weight of the trap $M_+$ is simply the probability $l_-/(l_+  + l_-)$ of reaching $M_+$ before $M_-$ in a flat landscape
\begin{eqnarray} 
P_{eff}(x, t) \sim \frac{l_+}{l_+ + l _ -} \delta(x+l _ -) + \frac{l_-}{l_+ + l _ -} \delta(x-l_+) 
\label{pefftrap} 
\end{eqnarray}
    This shows that the dynamics always remains out-of-equilibrium :  the weights of the two traps are not given by Boltzmann factors, they do not even depend on the energies of the traps, but only on their distances to the origin.

In the rescaled variable $X=x/\xi(t)$ where $\xi(t) \sim t^{\mu/(1+\mu)}$
represents the anomalous diffusion length scale,
 the average of the diffusion front (\ref{pefftrap}) 
over the samples reads \cite{trapsymmetric}
\begin{eqnarray} 
g_{\mu}(X) = e^{ - \vert X \vert} \int_0^{+\infty} du e^{-u}  \frac{u}{\vert X \vert +u} +O(\mu) 
\end{eqnarray} 

The localization parameters, which represents the averages over the samples of the probabilities to find $k$ independent particles on the same site, are given by \cite{trapsymmetric}
\begin{eqnarray} 
Y_k (\mu) \equiv \lim_{t \to \infty} \overline{ \sum_{n=0}^{+\infty} P^k(n, t\vert 0,0)} = \frac{2}{(k+1)} +O(\mu) 
\label{resyksummary} 
\end{eqnarray} 
This result is in agreement with the numerical simulations of Bertin and Bouchaud \cite{bertinjp} who have obtained $Y_2 \to 2/3$ and $Y_3 \to 1/2$ in the limit $\mu \to 0$.

 The thermal width reads
 \begin{eqnarray} 
c_2(\mu) \equiv \lim_{t \to \infty} \overline{\frac{<n^2>-<n>^2} { \xi^2(t)}} = 1+O(\mu) \end{eqnarray} 
and more generally, the others thermal cumulants can be derived from the generating function 
\begin{eqnarray} 
Z_{\mu}(s) \equiv \overline{ \ln < e^{-s \frac{n}{\xi(t)}} >} = \int_0^{+\infty} d \lambda e^{ - \lambda} \lambda \left(\frac{s \lambda}{2} \coth \frac{s \lambda}{2}-1 \right) +O(\mu) \end{eqnarray}

The two-particle correlation function reads
\begin{eqnarray}
C(l,t) && \equiv \overline{  \sum_{n=0}^{+\infty} \sum_{m=0}^{+\infty}
P(n,t\vert 0,0) P(m,t\vert 0,0) \delta_{l,\vert n-m \vert}}  \opsimeq_{t \to \infty} Y_2(\mu) \delta_{l,0}
+ \frac{1}{ \xi(t)} { C}_{\mu} \left( \frac{l}{\xi(t)} \right)
\nonumber \label{correform}
 \end{eqnarray}
The weight of the $\delta$ peak at the origin correspond
as it should to the localization  parameter $Y_2
=2/3 +O(\mu)$ (\ref{resyksummary}),
whereas the second term involves the following scaling function 
\begin{eqnarray}
 { C}_{\mu} (\lambda)= e^{-\lambda} \frac{\lambda}{3}   +O(\mu) 
\label{correlong} 
\end{eqnarray}

Finally, one can also obtain results for the two aging correlations
\cite{trapsymmetric},
namely for the probability $\Pi(t+t_w,t_w)$ of no jump during the time interval
 $[t_w,t_w+t]$, that presents a sub-aging scaling form in
 $t/t_w^{1/(1+\mu)}$, and for 
the probability $C(t+t_w,t_w)$ of being at time $(t+t_w)$ 
in the trap where it was at time $t_w$, that presents an aging scaling in
 $t/t_w$.

The strong disorder approach can also be used to obtain results
on the non-linear response in the aging regime \cite{trapreponse}.
  
  \section{ Wetting and other depinning transitions
 : random critical properties }

\label{wettingPS}

  \subsection{General introduction on random critical points  }
  
  \subsubsection{ Harris criterion to determine disorder relevance near  2d pure critical points}

The stability of pure critical points with respect
to weak bond disorder is governed by the Harris criterion \cite{harris} :
near a second order phase transition in dimension $d$,
the bond disorder is irrelevant 
if the correlation length exponent $\nu_P \equiv \nu_{pure} > 2/d$.
On the contrary if $\nu_{P} < 2/d$, disorder is relevant
and drives the system towards a random fixed point
characterized by new critical exponents.

A simple argument to understand Harris criterion is the following.
The pure system at a temperature $T \neq T_c$ is characterized by
a correlation length $\xi(T) \sim t^{-\nu_P}$, where $t= \vert T_c-T \vert$
represents the distance to criticality, and $\nu_P$ the correlation length exponent.
The pure system can be divided into
nearly independent subsamples of volume $V \sim \xi^d(T) \sim t^{-d \nu_P}$.
In the presence of an additional weak bond disorder, the averaged bond value $(1/V) \sum_{i \in V} J_i$
seen in a volume $V$ will present fluctuations of order $1/\sqrt{V}$.
So the fluctuations of critical temperatures among the volumes of size $V$
will be of order $\Delta T_c(V) \sim 1/\sqrt{V} \sim \xi^{-d/2}(T) $.
Disorder will be irrelevant if these fluctuations $\Delta T_c(V)
 \sim t^{d \nu_P/2}$ becomes negligeable
with respect to $t=\vert T_c-T \vert$ in the limit $t \to 0$ where the critical point is approached.

 \subsubsection{ General bound $\nu_{FS} \geq 2/d$ for random systems, and
the possible existence of two distinct exponents $\nu$ }
 
There exists a general bound for the finite-size correlation length exponent
 $\nu_{FS} \geq 2/d$ in disordered systems \cite{chayes}, 
which essentially means
 that a random critical point should itself be stable with
 respect to the addition of disorder,
 as in the Harris criterion argument given above.
 However, this general bound has to be understood with the subtleties 
explained in \cite{chayes}.
 In so-called `conventional' random critical points, 
there is a single correlation length exponent
 $\nu=\nu_{FS}$ and this single exponent is expected to satisfy the bound.
 However, there are also `unconventional' random critical points, where
 there are two different correlation length exponents ! 
In this case, the `intrinsic ' correlation
 exponent $\nu_{intrinsic}$ can be less than $2/d$,
 whereas the bound holds for the finite-size exponent $\nu_{FS} \geq 2/d$.
 The most well-understood example of the existence 
of two different correlation length exponents is
 the random transverse field Ising chain 
(this quantum 1D model is equivalent to
 the 2D classical Ising model with columnar disorder introduced by McCoy and Wu \cite{mccoywu}),
 which has been studied in great details by D. Fisher via a strong disorder renormalization approach
 \cite{danielrtfic} : the exponent $\tilde \nu=1$ governs the decay of the typical correlation 
 $ \overline{ \ln C(r) } \sim - r/ {\tilde \xi}$, whereas $\nu=2$ governs the decay of the averaged correlation 
 $ \ln ( \overline{  C(r)} ) \sim - r/ { \xi}$. Exactly at criticality, the typical and averaged correlations
 are also very different, since the typical correlation decays as $C_{typ} (r) \sim e^{- w { \sqrt r}}$,
 where $w$ is a random variable of order 1, whereas the averaged correlation
is dominated by rare events and decays algebraically
 $  \overline{  C(r)}  \sim 1/ r^{ (3-\sqrt 5)/2}$.
 Other critical points with two different correlation length exponents are discussed in \cite{singh,paz1,fisher2nu,bolech,myers}.

  \subsubsection{ Lack of self-averaging at random critical points}

 In disordered systems, the densities of extensive thermodynamic observables are self-averaging off-criticality,
because  the finiteness of the correlation length $\xi$
allows to divide a large sample into independent large
sub-samples.
At criticality however, this 'subdivision' argument breaks down
because of the divergence of $\xi$ at $T_c$, and a  
 lack of self-averaging has been found at criticality
 whenever disorder is relevant
\cite{domany95,AH,domany}. More precisely, for a given observable $X$,
it is convenient to define its normalized width as
\begin{equation}
\label{defratiodomany}
R_X(T,L) \equiv \frac{ \overline { X_i^2(T,L)} - ( \overline{X_i(T,L)})^2
}{ ( \overline{X_i(T,L)})^2 } 
\end{equation}
In terms of the correlation length  $\xi(T)$, the following behaviour of
$R_X(T,L)$ is expected \cite{AH,domany}  

(i) off criticality, when $L \gg \xi(T)$, 
the system can be divided into nearly independent sub-samples
and this leads to `Strong Self-Averaging' 
\begin{equation}
R_X(T,L) \sim \frac{1}{ L^d} \ \ \hbox{ off  criticality  for 
 $L \gg \xi(T)$ } 
\end{equation}

(ii) in the critical region, when $L \ll \xi(T)$, 
the system cannot be divided anymore into nearly independent sub-samples.
In particular at $T_c(\infty)$ where $\xi=\infty$,
one can have either `Weak Self-Averaging' 
\begin{equation}
\label{weaksa}
R_X(T_c(\infty),L) \sim \frac{1}{ L^{d-\frac{2}{\nu_P}}} 
 \ \ \hbox{ for  irrelevant
disorder ($\nu_{P} > 2/d$)  } 
\end{equation}
or `No Self-Averaging'
\begin{equation}
\label{nosa}
R_X(T_c(\infty),L) \sim Cst \ \ \hbox{ for   random  critical points  }
\end{equation}
To understand the origin of this lack of self-averaging, it is useful
to introduce the notion of sample-dependent pseudo-critical temperatures,
as we now explain.

  \subsubsection{ Distribution of pseudo-critical temperatures}

Important progresses have been made recently in the understanding of
finite size properties of random 
critical points \cite{domany95,AH,paz1,domany,AHW,paz2} .
To each disordered sample $(i)$ of size $L$, one should first associate
a pseudo-critical temperature $T_c(i,L)$  \cite{domany95,paz1,domany,paz2}. 
The disorder averaged pseudo-critical critical
temperature $T_c^{av}(L) \equiv \overline{T_c(i,L)}$
satisfies 
\begin{equation}
T_c^{av}(L)- T_c(\infty) \sim L^{-1/\nu_{R}}
\label{meantc}
\end{equation}
where $\nu_R$ is the correlation length
exponent. Eq. (\ref{meantc}) generalizes the analogous relation for
pure systems 
\begin{equation}
\label{puretc}
T_c^{pure}(L) - T_c(\infty) \sim L^{-1/\nu_{P}}
\end{equation}
The nature of the disordered critical point then depends on the
width $\Delta T_c(L)$ of the distribution of the pseudo-critical
temperatures $T_c(i,L)$
\begin{equation}
\Delta T_c(L) \equiv \sqrt{Var [T_c(i,L)]}
=\sqrt{\overline{T_c^2(i,L)}-\left(\overline{T_c(i,L)}\right)^2}
\end{equation}
When the disorder is irrelevant, the fluctuations
of these pseudo-critical temperatures obey
the scaling of a central limit theorem as in the Harris argument : 
\begin{equation}
\Delta T_c(L) \sim L^{-d/2} \ \ \hbox{ for  irrelevant  disorder }
\label{deltatcirrelevant}
\end{equation}
This behaviour was first believed to hold in general \cite{domany95,paz1}, 
but was later shown to be wrong in the case of random fixed points.
In this case, it was argued \cite{AH,domany} that
eq. (\ref{deltatcirrelevant}) should be replaced by
\begin{equation}
\Delta T_c(L) \sim L^{-1/\nu_{R}} \ \ \hbox{ for  random critical points }
\label{deltatcrelevant}
\end{equation}
i.e. the scaling is the same as the $L$-dependent shift of the averaged 
pseudo-critical temperature (Eq. \ref{meantc}).
The fact that these two temperature scales remain the same
is then an essential property of random fixed points
that leads to the lack of self-averaging at criticality.

Up to now, to our knowledge, 
the distribution of $T_c(i,L)$ or of another sample-dependent
 critical parameter has been studied for
 various disordered spin models 
\cite{paz1,domany,paz2,igloipottsq}, 
for elastic lines in random media \cite{bolech},
for Poland-Scheraga models
\cite{PS2005} and for the selective interface model \cite{interface2005}.
We will describe below the case of Poland-Scheraga model,
since the variation of one parameter in the Poland-Scheraga model
 allows to study disorder effects on both first order and second order transitions,
with either marginal or relevant disorder.

  \subsubsection{ Finite-size scaling in disordered systems }

In pure systems, the finite-size scaling theory
 relates the critical exponents of the phase transition in the thermodynamic limit
to finite-size effects that can be measured in numerical simulations \cite{cardyFSS}.
In short, this theory says that the only important variable is
the ratio between the size $L$ of the finite system and
the correlation length that diverges at the critical point $\xi(T) \sim \vert  T-T_c \vert^{-\nu}$.
So the data $X_L(T)$ for various sizes $L$ should be analysed
in terms of the appropriate rescaled variable $\tau = (T-T_c) L^{1/\nu}$
to obtain a master curve of the form $ L^y X_L(T) =\phi( \tau)$,
where $y$ is the exponent describing the decay the observable
$X$ exactly at criticality $X_L(T_c) \sim 1/L^y$.
Note that using the scale-dependent $T_c(L)$ instead of the thermodynamic $T_c=T_c(\infty)$
is completely equivalent, since it 
corresponds to a simple translation $\tilde \tau= (T-T_c(L)) L^{1/\nu}= \tau +a $, as a consequence of (\ref{puretc}).

In random systems, one has instead data $X^{(i)}_L(T)$ measured at temperature $T$
for various disordered samples $(i)$ of size $L$, and the question is :
what is the best way to analyse these data?
The usual procedure consists in averaging over the samples $(i)$ at fixed $(T,L)$
to apply the pure procedure to these disorder averaged quantities :
one tries to find a master curve $ L^y \overline{ X^{(i)}_L(T) }  =\phi( \tau)$
in terms of the variable $\tau = (T-T_c) L^{1/\nu}$.
However, this procedure leads to extremely large sample-to-sample fluctuations
in the critical region, as a consequence of the width of the
distribution of pseudo-critical temperatures : at a given temperature $T$, the samples
having their pseudo-critical temperature $T_c(i,L)>T$
are effectively in the low temperature phase, whereas the samples having  
$T_c(i,L)<T$
are effectively in the high temperature phase. This mixing of samples
in the critical regions makes it very difficult to obtain
clean results on critical exponents.

To avoid these difficulties,
the following alternative procedure has been proposed \cite{paz1,domany,paz2} :
the data for a given sample $(i)$ should be analysed in terms
of the rescaling variable $ \tau_{new}= (T-T_c(i,L)) L^{1/\nu}$ with respect to
its own pseudo-critical temperature $T_c(i,L)$. 
Since $T_c(i,L)=T_c^{av}(L)+u_i \Delta T_c(L)$, where $u_i$ is a random variable
of order one, and where the mean $T_c^{av}(L) $ and the variance
$\Delta T_c(L)$
follow the respective behaviours (\ref{meantc}) and (\ref{deltatcrelevant}), it is clear that the two procedures
are not equivalent, since $\tau_{new}=\tau+a+u_i$ is not a simple translation of the constant $a$
as in the pure case, because the random variable $u_i$ remains present for arbitrary $L$.
This new way of analysing the data allows to reduce very significantly
the sample-to-sample fluctuations, as shown for spin models \cite{paz1,domany,paz2} ,
for the non-equilibrium depinning transition of
 elastic lines in random media \cite{olaf}, and for disordered polymer models \cite{PS2005}.

   \subsection{Application to Poland-Scheraga model with various loop exponent $c$  }

\subsubsection{  Poland-Scheraga model of DNA denaturation  } 

The Poland-Scheraga model of DNA denaturation \cite{Pol_Scher} describes 
the configuration of the two complementary chains
as an alternance of bound segments and open loops.
Each loop of length $l$ has a polymeric entropic weight ${\cal N} (l) \sim \mu^l /l^c $,
whereas each contact at position $\alpha$ has for Boltzmann weight $e^{- \beta \epsilon_{\alpha}}$.
The partition function $Z(\alpha;1)$ with bound-bound ends at $(\alpha,1)$
thus satisfies the 
simple recursion relation
\begin{equation}
Z(\alpha;1)=  e^{-\beta \epsilon_{\alpha} }  
  \sum_{\alpha'=1}^{\alpha-1}   {\cal N}(\alpha-\alpha') Z(\alpha';1)
\label{recursion}
\end{equation}

The wetting model presented in the Introduction
(\ref{zwetting}) corresponds to a
Poland-Scheraga model with loop exponent $c=3/2$
(this exponent comes from the first return distribution of a 1D random walk).

For DNA denaturation, the question is what is the appropriate value of the loop exponent $c$?
Gaussian loops in $d=3$ dimensions are
characterized by $c=d/2=3/2$. The role of self avoidance
within a loop was taken into account by Fisher \cite{Fisher}, and
yields the bigger value $c=d \nu_{SAW} \sim 1.76$, where $\nu_{SAW}$ is the SAW radius of gyration
exponent in $d=3$. More recently, Kafri et al. \cite{Ka_Mu_Pe1,Ka_Mu_Pe2}
pointed out that the inclusion of the self avoidance of the loop with
the rest of the chain further increased $c$ to a value $c>2$.
 The transition in the pure
model should then be discontinuous, as previously found by
\cite{Barbara1} in Monte Carlo simulations of SAW 's. The value $c
\simeq 2.11$ was in turn measured in three dimensional Monte Carlo
simulations \cite{Carlon, Baiesi1, Baiesi2}.

\subsubsection{  Pure critical properties and disorder relevance : role of exponent $c$ } 

In the pure case $\epsilon_{\alpha}=\epsilon_0$, the model is of course exactly solvable,
and it turns out that the critical properties are determined by the value of the loop exponent $c$ :
for $c>2$, the transition is first order with exponent $\nu_P=1$,
whereas for $1<c<2$ the transition is second order with exponent $\nu_P=1/(c-1)$.
The Harris criterion thus yields that disorder is irrelevant for $c<3/2$,
marginal for $c=3/2$,
 and relevant for $c>3/2$.
 
 As explained above, the case $c=2.15$ where the pure transition
is first order is of interest for DNA denaturation. The effect of
disorder on this transition has been recently debated
\cite{Barbara2,adn2005,Gia_Ton}.

 The marginal case
$c=\frac{3}{2}$ has been debated for a long time
\cite{FLNO,Der_Hak_Van,Bhat_Muk,Ka_La,Cu_Hwa,Ta_Cha,wetting2005} 
and is of special interest since it corresponds to two-dimensional wetting
as already mentioned. On the analytical side, efforts have focused on the
small disorder limit : Ref \cite{FLNO} finds a marginally
irrelevant disorder where
 the quenched critical properties are the same as in the pure case,
 up to subdominant logarithmic corrections. Other
studies have concluded that that the disorder is marginally relevant
\cite{Der_Hak_Van,Bhat_Muk,Ka_La}. On
the numerical side, the same debate on the disorder 
relevance took place. The numerical studies of Ref. \cite{FLNO}
and Ref. \cite{Cu_Hwa} have concluded that the
critical behaviour was indistinguishable from the pure transition. On
the other hand, the numerical study of \cite{Der_Hak_Van}
 pointed towards a negative specific heat exponent
($\alpha<0$), and finally Ref. \cite{Ta_Cha}
has been interpreted as an essential singularity in the specific heat,
that formally corresponds to an exponent $\alpha=-\infty $.

  \subsubsection{  Distribution of pseudo-critical temperatures in Poland-Scheraga models }

In \cite{PS2005}, the distribution of pseudo-critical temperatures 
disordered Poland-Scheraga models with different 
loop exponents $c$, corresponding to
either a pure second order transition with respectively
marginal/relevant disorder according to the Harris criterion,
or to a pure first-order transition.
Two different definitions
for the pseudo-critical temperature have been considered, one  based on the free-energy,
and the other based on
a sample-replication procedure.
Both definitions
actually yield similar results for critical exponents and scaling
distributions, even though they give different values for a given sample.
This shows that the conclusions that can be obtained from such studies
do not depend on the precise definition of the pseudo-critical temperature.

 In all cases ($c=1.5$, $1.75$ and $2.15$),
and for both definitions of the pseudo-critical temperature
the distributions of pseudo-critical temperatures
were found to follow the scaling form
\begin{equation}
P_L(T_c(i,L)) \simeq  \frac{1}{ \Delta T_c(L)} \  g \left( x= \frac{
T_c(i,L) -T_c^{av}(L)}{ \Delta T_c(L) }  \right) 
\label{rescalinghistotc}
\end{equation}
where the scaling distribution $g(x)$ is simply Gaussian
\begin{equation}
g(x)=  \frac{1}{ \sqrt{2 \pi} } e^{- x^2/2}
\end{equation}
Note however that this Gaussian distribution is not generic
but seems specific to these Poland-Scheraga models, since in 
the selective interface\cite{interface2005}, the corresponding scaling distribution
was found to be very asymmetric.
The rescaling (\ref{rescalinghistotc}) means that the important scalings
of the pseudo-critical temperatures distribution
are the behaviours of its average $T_c^{av}(L)$ 
and width $\Delta T_c(L)$  as $L$ varies.

 For the marginal case $c=3/2$ corresponding to two-dimensional wetting, 
both the width $\Delta T_c(L)$ and the shift
$[T_c(\infty)-T_c^{av}(L)]$  are found to decay as $L^{-\nu_{random}}$, where the exponent is
very close to the pure exponent ($\nu_{random} \sim 2=\nu_{pure}$) but disorder is 
nevertheless relevant since it 
leads to non self-averaging at criticality.
For relevant disorder
$c=1.75$, the width $\Delta T_c(L)$ and the shift
$[T_c(\infty)-T_c^{av}(L)]$ decay with the same new exponent
$L^{-1/\nu_{random}}$ (where $\nu_{random} \sim 2.7 > 2 > \nu_{pure}$) and
there is again no self-averaging at criticality. Finally for the value
$c=2.15$, of interest in the context of DNA denaturation, the
transition is first-order in the pure case. In the presence of
disorder, the width $\Delta T_c(L) \sim L^{-1/2}$ 
dominates over the shift $[T_c(\infty)-T_c^{av}(L)] \sim L^{-1}$,
i.e. there are two correlation length exponents $\nu=2$ and $\tilde
\nu=1$ that govern respectively the averaged/typical loop distribution.
This is reminiscent of what happens at strong disorder
fixed points \cite{danielrtfic,revue}, and perhaps indicates that
some appropriate strong disorder RG description
of this random critical point should be possible.

  \section{ Conclusion}

We have presented some recent results concerning
the effects of quenched disorder on random walks and polymer models. 
For random walk models, we have explained how the Golosov localization
phenomenon allows to construct a strong disorder approach to obtain
 detailed explicit results
on the anomalous diffusion behaviours and on aging properties.
For disordered polymer models presenting depinning transitions,
we have explained how the recent progresses 
in the general theory of random critical points
are very useful to understand the critical properties.

To finish, let us mention some closely related works.
In the field of random walks, we have only described here
the simplest models involving a single random walk in 
a disordered medium,
but models involving many interacting
random walks in the presence of quenched disorder
can be also studied via the strong disorder approach \cite{revue},
in particular reaction diffusion models \cite{rgrfim,rgreadiff},
asymmetric exclusion processes \cite{igloiexclusion},
contact processes \cite{igloicontact} and zero-range processes
 \cite{igloizerorange}.
In the field of phase transitions of disordered polymer models,
we have restricted our presentation here to the case of
bidimensional wetting and of
 Poland-Scheraga model of DNA denaturation, but the idea to study
the distribution of pseudo-critical temperatures
allows to clarify the critical properties of other equilibrium transitions,
such as the selective interface model \cite{interface2005} 
and the directed polymer transition in $d=3$ \cite{future}
mentioned in the introduction (\ref{directed}).
This framework is also very useful 
for the non-equilibrium depinning transition of
 elastic lines in random media, where the notion of pseudo-critical
temperature is replaced by the notion of sample-dependent
pseudo-critical force \cite{bolech,olaf}.

 \begin{acknowledgments}

It is a pleasure to thank D.S. Fisher and P. Le Doussal for collaboration
on random walks in random media 
\cite{rgsinaishort,rgsinailong,rgrfim}, P. Le Doussal for 
the continuation of this initial collaboration in various directions 
\cite{rgreadiff,locgolosov,energysinai,rgtoy},
F. Igloi for cowriting the review article on
the strong disorder approach of random systems
\cite{revue},
and T. Garel for past and present collaboration on polymer models
with quenched disorder 
  \cite{directed2004,wetting2005,adn2005,PS2005,interface2005,future}.

I also wish to thank for interesting discussions
the mathematicians  A. Bovier,  J. Cerny, A. Faggionato, Y. Hu, Z. Shi
for the random random walks part, and 
T. Bodineau, G. Giacomin,  F. Toninelli
for the disordered polymers part.

\end{acknowledgments}

\end{document}